\documentstyle[12pt,physrev]{article}

\begin{document}

\begin{center}

{\Large {\bf Self weakening of the tensor interaction in a nucleus}}

\vspace{0.3in}

L. Zamick $^{a)}$, D. C. Zheng $^{b)}$, and M. Fayache $^{a)}$\\

\end{center}

\vspace{0.2in}

\begin{small}

\noindent
$^{a)}${\it Department of Physics and Astronomy, Rutgers University,
	Piscataway, New Jersey 08855}

\noindent
$^{b)}${\it Department of Physics, University of Arizona,
	Tucson, Arizona 85721}

\end{small}

\thispagestyle{empty}

\vspace{0.5in}

\begin{abstract}
We examine several ``landmarks'' for the effects of the tensor
interaction on the properties of light nuclei. These were
usually discussed in the context of small-space shell-model calculations.
We show, using $G$ matrices derived from a realistic nucleon-nucleon
potential, that when the model space is small (e.g., $0\hbar\omega$),
these effects are overestimated, indicating that the tensor
interaction is too strong. However, when larger spaces are used,
there is a diminishing of these effects, which, in general, leads
to better agreement with experiment.
\end{abstract}

\pagebreak

\section{Introduction}
The tensor interaction in a nucleus is a rather elusive beast.
There are no first-order contributions to the binding energy of a closed
$LS$ nucleus like $^4\mbox{He}$, $^{16}\mbox{O}$, or $^{40}\mbox{Ca}$
due to the tensor interaction. Likewise, there are no first-order
contributions to the single-particle energies for a closed $LS$ shell
plus or minus one nucleon. There are, however, important second-order
contributions.

To see the effects of the tensor interaction in first order, the simplest
thing to do is to go to systems with {\it two} quasi-particles.
To this end, we will consider the following cases:

\vspace{0.1in}
\begin{small}

\hspace{0.5in} a) 1p-1h system: the $J^{\pi}$=$0^-$, $T$=0 and $T$=1
	          states in $^{16}\mbox{O}$,

\hspace{0.5in} b) 2-hole system: the beta decay of
		  $^{14}\mbox{C}$ to $^{14}\mbox{N}$,  and

\hspace{0.5in} c) 2-particle system: the E2 and M1 moments of $^{6}\mbox{Li}$.

\end{small}

\vspace{0.1in}

For the above cases, we perform shell-model calculations first in a small
space ($0\hbar\omega$) and then in larger spaces in which $2\hbar\omega$
and sometimes even higher excitations are included.

We employ Brueckner reaction matrices $G$ \cite{bruc} calculated according to:
\begin{equation}
G(E_s) = v_{12} + v_{12} \frac{Q}{E_s-(h_1+h_2+v_{12})} v_{12},
					\label{G}
\end{equation}
using the method introduced in Ref.\cite{bhm}.
In the above equation, $v$ is the bare $NN$ potential for which we adopt a
new Nijmegen local $NN$ interaction (NijmII) \cite{nijm};
$h=t+u$ is the one-body Hamiltonian with $u$ chosen to be
the harmonic-oscillator (HO) potential $u(r)=\frac{1}{2}m\omega^2r^2$;
$E_s$ is the starting energy,
which, for an initial two-particle state $|12\rangle$
in the ladder diagrams, is taken to be
\begin{equation}
E_s = \epsilon_1 + \epsilon_2 + \Delta\, ,   \label{Es}
\end{equation}
where $\epsilon$'s are the HO single-particle energies,
\begin{equation}
\epsilon_i = \left( 2n_i+l_i+\frac{3}{2}\right)\hbar\omega
	\equiv \left( N_i+\frac{3}{2}\right)\hbar\omega \, ,
\end{equation}
thus $(\epsilon_1 + \epsilon_2)$ is the unperturbed energy of the
initial two-particle state $|12\rangle$ in the ladder diagrams.
The parameter $\Delta$ in Eq.(\ref{Es}) can be thought of as
the interaction energy between the two particles.
The reader is referred to Ref.\cite{mfd} for a partial justification for
using the prescription in Eq.(\ref{Es}) for the starting energy.
Note that such a state-dependent choice for $E_s$ leads to
a non-hermitian $G$ matrix but the non-hermiticity is found to be
small and we here use an average of $G$ and its hermitian conjugate,
$\frac{1}{2}(G+G^+)$, as our effective interaction $v^{\rm eff}$.
For all the nuclei under consideration,  we fix the basis parameter
$\hbar\omega$ at 14 MeV and the starting-energy
parameter $\Delta$ at -50 MeV.

The Pauli operator $Q$ in Eq.(\ref{G}) is defined to prevent
the two nucleons from scattering into the intermediate states which are
either occupied (therefore Pauli forbidden) or inside the model space
(to avoid double counting).
It is therefore related to the choice of the model space.
As we increase the size of the model space, we enlarge the $Q$=0
region to avoid double counting. We show in Fig.1, as an example,
our definition of $Q$ which is used to calculate the $G$ matrix
for a full $3\hbar\omega$-space calculation of $^{16}\mbox{O}$.
We cut off the ``wings'' at the edge of the $N$=8 major shell.
We emphasize that in this work we have used
different definitions of $Q$ for different model spaces and/or nuclei.

The matrix diagonalization are performed for the shell-model Hamiltonian
\begin{equation}
H_{\rm SM} = \left( \sum_{i=1}^A t_i -T_{\rm c.m.}\right)
	+ \sum_{i<j}^A v^{\rm eff}_{ij}
	+ \lambda (H_{\rm c.m.} - \frac{3}{2}\hbar\omega),
				\label{hsm}
\end{equation}
where $T_{\rm c.m.}$ is the center-of-mass (c.m.) kinetic energy and
the last term (with $\lambda \gg 1$) is added to remove the
spurious effects of the c.m.~motion from the low-lying states.
We have not included the Coulomb interaction.
It should be pointed out that our calculations involve no
phenomenological single-particle energies. These are implicitly
generated from the two-body $G$ matrix elements as well as the one-body
kinetic energy in the matrix diagonalization.

\section{The 1p-1h Case: Isospin Splitting of $J^{\pi}$=$0^-$
 States in $^{16}\mbox{O}$}

We previously considered in some detail the isospin splitting of
$J^{\pi}$=$0^-$ states in $^{16}\mbox{O}$ \cite{zzfm}. We will here
take the opportunity to discuss a few points. It was shown by
Blomquist and Molinari \cite{bm} and Millener and Kurath \cite{mk}
that without the tensor interaction, the energy difference $\Delta E$
between the $J^{\pi}$=$0_1^-$, $T$=1 and $J^{\pi}$=$0_1^-$, $T$=0 states
in $^{16}\mbox{O}$ would be very small. Experimentally, the
$J^{\pi}$=$0_1^-$, $T$=0 state is at 10.952 MeV and the $J^{\pi}$=$0_1^-$,
$T$=1 is at 12.797 MeV so that the value of $\Delta E$ is 1.845 MeV.
The results for $\Delta E$ using various model spaces, using the
$G$ matrices derived from the NijmII potential, are given in Table I.

In the $1\hbar\omega$ 1p-1h space, in which the dominant configuration is
$(1s_{1/2} 0p_{1/2})^{J^{\pi}=0^-}$ with small admixture of
$(0d_{3/2} 0p_{3/2})^{J^{\pi}=0^-}$, the value of $\Delta E$ is too large,
i.e., 2.809 MeV as compared to the experimental value of 1.845 MeV.
Thus from this case one might deduce that the effective tensor interaction
is too strong. The situation is  not improved much when $3\hbar\omega$
3p-3h configurations are included in the matrix diagonalization,
i.e., when, besides the $1\hbar\omega$ 1p-1h already there,
we allow two additional nucleons to be excited from
the $0p$ shell to the $1s$-$0d$ shell. The splitting $\Delta E$ changes
from 2.809 MeV to 2.678 MeV, see Table I.

However, when $3\hbar\omega$ 2p-2h admixtures [i.e.,
$(0s)^{-1}(0p)^{-1} (1s0d)^2$ and $(0p)^{-2} (1s0d)^1(1p0f)^1$]
are introduced, the situation
improves dramatically: $\Delta E$ goes down to 1.641 MeV. This
number is in better agreement with experiment but we have an overshoot.
When we furthermore include $3\hbar\omega$ 1p-1h configurations
[i.e., $(0s)^{-1}(1p)^1$ and $(0p)^{-1} (2s1d0g)^1$] to make
our model space complete for a full $(1+3)\hbar\omega$ calculation,
the splitting becomes 1.931 MeV, in very good agreement with to experiment.

This is one example of the ``self-weakening'' mechanism for the
effective tensor interaction. The 2p-2h diagrams which
contribute to $\Delta E$
in second-order perturbation theory are shown in Fig.2. We classify
them as particle-hole (or bubble) (Fig.2a),
hole-hole (Fig.2b), and particle-particle (Fig.2c) diagrams.
In order to show how different parts of the interaction contribute
to the isospin splitting $\Delta E$, we present the perturbation-theory
results in Table II for a schematic interaction which was introduced
in Ref.\cite{zzann} for an easy control of the strengths of
the spin-orbit and tensor interactions:
\begin{equation}
V = V_c + x V_{s.o.} + yV_t \; ,
\end{equation}
where $c$=central, $s.o.$=spin-orbit, $t$=tensor.
For $x$=1 and $y$=1, the matrix elements of this interaction
approximately resemble $G$ matrix elements derived from
a realistic $NN$ potential like Bonn A \cite{bonn}.
By setting $x$=0 (1),
we switch the spin-orbit interaction off (on); by setting $y$=0 (1),
we switch the tensor interaction off (on).

As noted in Ref.\cite{zzfm}, the particle-hole diagram (Fig.2a)
is the most important for getting a large, negative contribution
for $\Delta E$, but this only occurs when the tensor interaction
is turned on. From Table II, we also note that the particle-particle
and hole-hole diagrams are of the same sign and they act against the
particle-hole diagrams. For $x$=1 and $y$=1, whereas the particle-hole
diagrams contribute an amount of -1.728 MeV to the splitting $\Delta E$,
the particle-particle and hole-hole diagrams together give a
contribution of +0.901 MeV, leading to a net result of
$\Delta E$=-0.826 MeV.

We wish to note, and this point has not been made before, that for
the particle-hole diagrams of Fig.2a,
the most important contribution comes from the case in which
the lower vertex involves a central interaction and the
upper vertex involves a tensor interaction. Indeed, when the $PH$ in Fig.2a
is equal to $(1s)(0s)^{-1}$, the lower vertex cannot involve a tensor
interaction at all because the matrix element for the lower vertex
involves only $l$=0 orbits. Thus Fig.2a is approximately linear in the
strength of the tensor interaction and it tends to act {\em against}
the first-order 1p-1h tensor interaction.

\section{The $A$=14 System (2 Holes): Allowed but Inhibited Beta Decay}

Another landmark example of the effects of the tensor interaction in a
nucleus is the famous $A$=14 beta decay:
$^{14}\mbox{C}(J^{\pi}\!=\!0^+, T\!=\!1)
\rightarrow ^{14}\mbox{N}(J^{\pi}\!=\!1^+, T\!=\!0)$.
The quantum numbers involved in this transition
are just right for an allowed Gamow-Teller
(GT) transition, but the transition is very strongly suppressed. The matrix
element $B({\rm GT})$ is essentially zero.

For the two holes in the $0p$ shell,
the wave functions of the initial and final states
in the $LS$ coupling can be written as \cite{zzann}
\begin{eqnarray}
\psi_i &=& C_i^S\; |{}^1S_0\rangle + C_i^P\; |{}^3P_0\rangle\; ,  \nonumber \\
\psi_f &=& C_f^S\; |{}^3S_1\rangle + C_f^P\; |{}^1P_1\rangle\;
				   + C_f^D\; |{}^3D_1\rangle \; .
\end{eqnarray}
It was shown by Inglis \cite{inglis} (see also Ref.\cite{wong})
that it is impossible to get $B({\rm GT})$=0 with the
above wave functions unless there is a tensor interaction present.
Since the GT operator $\sum \sigma_{\mu}(i) t_+(i)$
cannot change the orbital angular momentum, one way (but not the only way)
of getting $B({\rm GT})$=0 would be to have $\psi_f = |^3 D_1\rangle$.
In general, the GT amplitude is [$B({\rm GT})=|A({\rm GT})|^2$]
\begin{equation}
A({\rm GT}) = \sqrt{6}\left( C_i^SC_f^S - \frac{1}{\sqrt{3}}C_i^PC_f^P\right).
\end{equation}
The fact that the one-body spin-orbit force for holes is minus
that for particles (whereas the hole-hole two-body interaction is
equal to the particle-particle two-body interaction) leads to a large
admixture of $|^3D_1\rangle$
in the $J^{\pi}$=$1^+$ ground state of $^{14}\mbox{N}$.

Are higher-shell effects important here? To answer this question,
we give in Table III the results for $B({\rm GT})$ along with a few
other observables, obtained in the $0\hbar\omega$ and
$2\hbar\omega$ calculations using the NijmII \cite{nijm} $G$ matrices.
[As we mentioned in section I, when going from the $0\hbar\omega$ space
to the $2\hbar\omega$ space, we have modified the Pauli operator $Q$
in the $G$ matrix equation (\ref{G}) to exclude the model-space states
from the intermediate spectrum.]
As a contrast we also show in Table III our results for a one-hole
system ($A$=15) for which there is much less sensitivity to
the $2\hbar\omega$ configuration mixing.

The results in Table III for $B({\rm GT})$ require some explanation.
The value of $B({\rm GT})$ for $A$=14 that we obtain in the $0\hbar\omega$
calculation is 3.967, much larger than zero. However, in terms of the tensor
interaction, this means that we have an {\em overshoot}. As discussed
in Ref.\cite{zzann}, the amplitude $A({\rm GT})$ is large in magnitude
when there is no tensor force.
As we turn on the tensor interaction and gradually increase its strength,
$A({\rm GT})$ decreases in magnitude, goes through zero and changes sign.
This happens well before we come to the full tensor strength.
When we further strengthen the tensor interaction towards the full scale,
$A({\rm GT})$ becomes large in magnitude again.
Therefore, the large result obtained for $B({\rm GT})$ in the
$0\hbar\omega$ space again indicates that the tensor interaction is
too strong.

When we go from the $0\hbar\omega$ space to the full $2\hbar\omega$
space, the value of $B({\rm GT})$ decreases by about 55\% from 3.967 to
1.795. This latter value is still far from satisfactory but it is much
closer to the experimental answer of nearly zero.

It should be pointed out that, unlike the example of the
isospin splitting between $J^{\pi}$=$0^-$ states in $^{16}\mbox{O}$ which
we discussed in section II, in the present case of the $A$=14 beta decay,
the spin-orbit interaction also plays an
important role. An alternate way of getting $B({\rm GT})$=0
is to keep the tensor interaction at its full strength but {\em increase}
the strength of the spin-orbit interaction. However, in a previous work
\cite{zzfm}, we have seen that the higher-shell effects do not have
a significant effect on the spin-orbit interaction as they do on the tensor
interaction.

The higher-shell admixtures also have a large effect on the
magnetic dipole (M1) moment $\mu$ of the ground state ($J^{\pi}$=$1^+$,
$T$=0) in $^{14}\mbox{N}$, which is $0.768\, \mu_N$ in the
$0\hbar\omega$ calculation and $0.554\, \mu_N$ in the
$2\hbar\omega$ calculation. These values are obtained
using the bare $g$ factors:
$g_l(p)$=1, $g_l(n)$=0, $g_s(p)$=5.586, $g_s(n)$=-3.826.
The experimental result is $0.404\, \mu_N$ \cite{ajz,data}.

For the $A$=15 system, we note that the higher-shell admixtures do not have a
significant effect on $B({\rm GT})$. We can understand this from the theorem
which says that there are no first-order corrections to $B({\rm GT})$, or
to the M1 moment for a system consisting of a closed $LS$ shell
plus or minus one nucleon. For the two-hole system, on the other hand,
the higher-shell effects can, in part, renormalize
the particle-particle (or hole-hole) interaction between
the two quasi-particles.

\section{The $A$=6 System (2 Particles):
The Magnetic Dipole Moment and Electric
Quadrupole Moments of $^6\mbox{Li}$}

One more landmark signature of the tensor interaction, although
one that is somehow not recognized by many people, is the fact
that the electric quadrupole (E2) moment of the ground state in $^6\mbox{Li}$
is {\em negative}.
To show this, we again use the schematic interaction previously
described. With the bare values of
$e_p$=1 and $e_n$=0, the quadrupole moment $Q$
of the $J^{\pi}$=$1^+$ ground state of $^6\mbox{Li}$
assumes the following values when we fix the spin-orbit interaction
at its full strength ($x$=1) and vary the strength ($y$)
of the tensor interaction:
\begin{eqnarray}
Q&=&\;\;\; 0.106\, e \,{\rm fm}^2  \hspace{0.2in}
		{\rm for} \hspace{0.1in} y=0\; , \nonumber \\
Q&=&-0.135\, e\, {\rm fm}^2  \hspace{0.2in}
		{\rm for} \hspace{0.1in} y=0.5\; , \nonumber \\
Q&=&-0.358\, e\, {\rm fm}^2  \hspace{0.2in}
		{\rm for} \hspace{0.1in} y=1\; . \nonumber
\end{eqnarray}
That is to say, when the tensor interaction is switched off, the
quadrupole moment is positive. As we increase the strength of the tensor
interaction, $Q$ decreases from being positive to being negative.

We also calculate the E2 moment $Q$ of $^6\mbox{Li}$
using the NijmII $G$ matrices.
The results from the shell-model diagonalization are shown in Table IV
for three model spaces: $0\hbar\omega$, $2\hbar\omega$, and $4\hbar\omega$.
We also give the results for the M1 moment $\mu$ obtained
using again the bare $g$ factors.

In the $0\hbar\omega$ space, the calculated value of $Q$ is
$-0.360 \,e\, {\rm fm}^2$, which is much more negative than the
experimental value of $-0.082\, e\, {\rm fm}^2$. Again, this could
be interpreted as being due to the fact that the effective
tensor interaction in this $0\hbar\omega$ space is too
strong. However, as we enlarge the model space, the magnitude of
$Q$ comes down. In the $2\hbar\omega$-space calculation, the value of $Q$ is
$-0.251\, e\,{\rm fm}^2$ and in the $4\hbar\omega$ space,
there is an overshoot: we obtain $-0.0085\, e\,{\rm fm}^2$.

For the M1 moment $\mu$, the experimental value is $0.822\, \mu_N$.
Because the ground state of $^6\mbox{Li}$ has isospin zero, this is an
isoscalar magnetic moment. The experimental value lies between the
value for the $jj$ limit ($0.627\, \mu_N$) and the value for the $LS$ limit
($0.880\, \mu_N$). Going from the $0\hbar\omega$ to the
$2\hbar\omega$ space, the calculated value of $\mu$ changes from
$0.866 \mu_N$ to $0.848 \mu_N$. We are still above the experimental value
even in the $4\hbar\omega$ calculation. The deviation is only
$0.024 \mu_N$ or 2.9\%. However, for the {\em isoscalar} moments, one
generally has a higher standard than for the isovector moments. The
experimental isoscalar moments lie much closer to the Schmidt limit than do
the isovector ones and they are less sensitive to configuration mixing.

We feel that $^6\mbox{Li}$ deserves further study. It is the most
elementary example of a system with two nucleons embedded in a nuclear
medium. The medium corrections can be calculated to a higher precision
than in heavier nuclei.

\section{Additional Remarks}
There are other approaches for dealing with light systems such as
cluster calculations for $^6\mbox{Li}$ performed by
Lehman {\it et al.} \cite{lehman},
Eskandarian {\it et al.} \cite{esk}, and
Schellingerhout {\it et al.} \cite{schel}.
As compared to our shell-model results, the above authors seem to get
better agreement for the isoscalar magnetic moment of the ground
state but the quadrupole moment of this state comes out positive.
One can argue that the cluster approach is more physical.
On the other hand, the Brueckner shell model is not self limiting.
With improved technology the calculations can always be
extended in a systematic manner.
This does not necessarily mean that perfect agreement with experiment
will be reached since the assumption that we can describe a nucleus
solely in terms of neutrons and protons interacting via a
two-body interaction may not be correct. But to see this requires
very high quality calculations.

Another point to be made is that, in contrast to this work where we
claim that many anomalies in {\em nuclear structure}
relating to the tensor interaction can be explained by simply
admixing higher-shell configurations, in the realm of
nucleon-nucleus scattering, there have been many works which
put forth the idea that the tensor interaction in the nuclear medium
must be considerably modified. These include experimental analyses of
Hintz {\it et al.} \cite{hintz} and Stephenson {\it et al.} \cite{step}.
They find that the polarization anomalies in proton-nucleus
scattering can be removed by adopting the theoretical
``universal scaling'' ideas of Brown and Rho \cite{brown} that
all mesons except the pion are less massive in the nuclear
medium.
Our next task will be to consider the fact that we do not
seem to need medium modifications of the tensor interaction
for nuclear structure, but we do need them for
nucleon-nucleus scattering.

This work was supported in part by a DOE grant
DE-FG05-86ER-40299 and an NSF grant PHY-9321668.

\pagebreak

\begin{small}

\noindent
{\bf Table I.} The isospin splitting $\Delta E$ of the lowest $0^-$ states
in $^{16}\mbox{O}$ obtained from shell-model diagonalizations in
various model spaces. The binding energy ($E_B$) for the ground state
and the excitation energy for the $0^-_1 T$ states ($T$ is the isospin) are
also listed. The Coulomb interaction is not included. (The
experimental binding energy listed in the table is Coulomb corrected.)
All energies are in MeV.

\begin{center}
\begin{tabular}{l|l|rrr|r} \hline\hline
\multicolumn{2}{c|}{Model space}
 & $E_B$ \hspace{0.1in} & $E(0^-_1,0)$ & $E(0^-_1, 1)$ & $\Delta E$\\ \hline
$0^+$: $0\hbar\omega$ & $0^-$: $1\hbar\omega$ 1p-1h
        &  118.933 & 15.943 & 18.752 & 2.809 \\
$0^+$: $(0+2)\hbar\omega$ & $0^-$: $1\hbar\omega$ 1p-1h
		+ $3\hbar\omega$ 3p-3h
	&  114.404 & 17.747 & 20.425 & 2.678 \\
$0^+$: $(0+2)\hbar\omega$ & $0^-$: $1\hbar\omega$ 1p-1h
		+ $3\hbar\omega$ 3p-3h + $3\hbar\omega$ 2p-2h
        &  124.692 & 15.946 & 17.587 & 1.641 \\
$0^+$: $(0+2)\hbar\omega$ & $0^-$: Full $(1+3)\hbar\omega$
	&  124.692 & 15.646 & 17.577 & 1.931 \\ \hline
\multicolumn{2}{c|}{Experiment}
      & $\sim 142\;\;$ & 10.952 & 12.797 & 1.845 \\ \hline\hline
\end{tabular}
\end{center}

\vspace*{0.5in}

\noindent
{\bf Table II.} The contributions (in MeV) of the two-particle, two-hole
perturbation-theory
diagrams of Fig.2 to the isospin splitting of the lowest $J=0^-$ states in
$^{16}\mbox{O}$.

\begin{center}
\begin{tabular}{c|c|ccc|c}\hline\hline
\multicolumn{2}{c|}{Interaction} & \multicolumn{3}{c|}{Diagram} & \\  \hline
Spin-orbit & Tensor & Particle-Hole & Particle-Particle
		    & Hole-Hole     & Total \\ \hline
Off & Off & 0.012 & 0.252 & 0.049 & 0.312 \\
On  & Off &-0.024 & 0.249 & 0.057 & 0.282 \\
Off & On  &-1.739 & 0.639 & 0.385 &-0.714 \\
On  & On  &-1.728 & 0.593 & 0.308 &-0.826\\ \hline\hline
\end{tabular}
\end{center}

\pagebreak

\noindent
{\bf Table III.} Properties of $A$=14 and $A$=15 nuclei from the
$0\hbar\omega$-space and $2\hbar\omega$-space shell-model matrix
diagonalizations. In the table, we also give the binding energy
$E_B(J^{\pi},T)$ for the ground state and the excitation energy
$E_x(J^{\pi},T)$ for the excited state involved.
Bare electromagnetic operators are used.

\begin{center}
\begin{tabular}{c|c|ccc}\hline\hline
$A$ & Observable & $0\hbar\omega$ & $2\hbar\omega$ & Exp't \\ \hline
14  & $B({\rm GT}) (0^+\, 1\rightarrow 1^+\, 0)$
			&  3.967  & 1.795 & $\sim 0$ \\
    & $B({\rm M1}) (0^+\, 1\rightarrow 1^+\, 0)(\mu_N^2)$
			&  9.737 &  4.998 &          \\
    & $\mu(1^+\, 0)(\mu_N)$
	                &  0.768 &  0.554 & 0.404 \\
    & $Q(1^+\, 0)(e\, {\rm fm}^2)$
		        &  1.236 &  2.151 & 1.56  \\
    & $E_B(1^+,0)$(MeV) & 82.928 & 89.521 & 104.64 \\
    & $E_x(0^+,1)$(MeV) &  2.142 &  1.836 & 2.313  \\ \hline
15  & $B({\rm GT}) (\frac{1}{2}^-\rightarrow \frac{1}{2}^-)$
			&  0.333  & 0.326 & 0.270\\
    & $\mu(^{15}\mbox{N})(\mu_N)$
			&  -0.264 &-0.277 & -0.283\\
    & $\mu(^{15}\mbox{O})(\mu_N)$
			&   0.638 & 0.655 & 0.719 \\
    & $E_B(\frac{1}{2}^+,\frac{1}{2})$(MeV)
			&  98.496 &104.778 & 115.476 \\
    & $E_x(\frac{3}{2}^+,\frac{1}{2})$(MeV)
			&   4.102 &  5.444 &   6.324 \\ \hline\hline
\end{tabular}
\end{center}

\vspace{0.5in}

\noindent
{\bf Table IV.} The results for the
electric quadrupole moment (in $e\, {\rm fm}^2$)
and magnetic dipole moment (in $\mu_N$) and the binding energy (in MeV)
of the ground state in $^6\mbox{Li}$ from
$0\hbar\omega$, $2\hbar\omega$, and $4\hbar\omega$ shell-model
calculations. Bare electromagnetic operators are used.
\begin{center}
\begin{tabular}{c|ccc}\hline\hline
Space  &  $Q$  & $\mu$  & $E_B$ \\ \hline
$0\hbar\omega$ &  -0.360  & 0.866  & 26.49 \\
$2\hbar\omega$ &  -0.251  & 0.848  & 27.58 \\
$4\hbar\omega$ &  -0.0085 & 0.846  & 30.03 \\
Exp't          &  -0.082  & 0.822  & 31.989 \\ \hline\hline
\end{tabular}
\end{center}

\end{small}

\pagebreak

\section*{Figure Captions}

\noindent
{\bf Fig.1} \hspace{0.1in} The Pauli operator $Q$ for a full
$3\hbar\omega$ calculation of $^{16}\mbox{O}$ in a model space consisting
of five major shells (s+p+sd+pf+sdg).
$Q$=0 for $N_1+N_2 \leq 5$ or $N_1 \leq 1$ or $N_2 \leq 1$.
Here $N_i=2n_i+l_i$ is the principal quantum number for
the harmonic oscillator single-particle state $|i\rangle$.
It is 0 for the lowest major shell ($0s$) and
1 for the $0p$ major shell, {\it etc.}
The wings extend out to include the nineth ($N$=8) major shell.

\vspace{1.0in}

\noindent
{\bf Fig.2} \hspace{0.1in} The 2p-2h admixtures to the 1p-1h
configuration: (a) particle-hole (bubble) diagrams,
(b) hole-hole diagram, and (c) particle-particle diagram.

\end{document}